\documentclass{article}
\usepackage{PRIMEarxiv}
\usepackage[colorlinks=true,linkcolor=red,urlcolor=blue,citecolor=blue,bookmarks=true]{hyperref}
\usepackage{lipsum}
\usepackage{amsfonts}
\usepackage{caption}
\usepackage{subcaption}
\usepackage{graphicx}
\usepackage{epstopdf}
\usepackage{amsmath}
\usepackage{amssymb}
\usepackage{cleveref}
\usepackage{amsopn}
\usepackage{setspace}
\usepackage{braket}
\usepackage{color}
\usepackage{algorithmicx}
\usepackage[ruled]{algorithm}
\usepackage{algpseudocode}
\usepackage{algpascal}
\usepackage{footnotebackref}
\usepackage{threeparttable, tablefootnote}
\usepackage[table]{xcolor}
\usepackage{amsthm}
\usepackage{latexsym}
\usepackage[mathscr]{euscript}

\DeclareMathAlphabet\mathbfcal{OMS}{cmsy}{b}{n}

\usepackage[utf8]{inputenc} 
\usepackage[T1]{fontenc}    
\usepackage{url}            
\usepackage{booktabs}        
\usepackage{nicefrac}       
\usepackage{microtype}      
\usepackage{fancyhdr}       
\graphicspath{{media/}}     

\providecommand{\JEL}[2]
{
  \small	
  \textbf{\textit{JEL Codes:}} #2
}

{\title{Machine Learning for Economics Research: \\ When What and How?\thanks{The opinions here are solely those of the authors and do not necessarily reflect those of the Bank of Canada. We thank Jacob Sharples for his assistance on the project. We also thank Andreas Joseph, Dave Campbell, Jonathan Chiu,  Narayan Bulusu, and Stenio Fernandes for their suggestions and comments on the article.}}} 

\author{
  Ajit Desai \\
  Bank of Canada \\
}

\begin{document}
\maketitle

\begin{abstract}
This article provides a curated review of selected papers published in prominent economics journals that use machine learning (ML) tools for research and policy analysis. The review focuses on three key questions: (1) when ML is used in economics, (2) what ML models are commonly preferred, and (3) how they are used for economic applications. The review highlights that ML is particularly used to process nontraditional and unstructured data, capture strong nonlinearity, and improve prediction accuracy. 
Deep learning models are suitable for nontraditional data, whereas ensemble learning models are preferred for traditional datasets.
While traditional econometric models may suffice for analyzing low-complexity data, the increasing complexity of economic data due to rapid digitalization and the growing literature suggests that ML is becoming an essential addition to the econometrician's toolbox.
\end{abstract}

\JEL J{A10, B23, C45, C55}

\onehalfspacing 
\section{Introduction}

The economy is becoming increasingly digital, and as a result, the size and complexity of economic data are growing rapidly. This presents both opportunities and challenges for analysts who want to process and interpret this data to gain insights into economic phenomena.
Machine learning (ML) which has emerged as a powerful tool for analyzing large and complex datasets across disciplines, has the potential to mitigate some of the challenges posed by the digitization of the economy for economics research and analysis.
The ability of ML models to effectively process large volumes of diverse data could allow us to build more complex models. 
As a result, the use of ML has expanded in economics research. 

\begin{figure}[h!]
    \centering
    \includegraphics[width=0.64\textwidth]{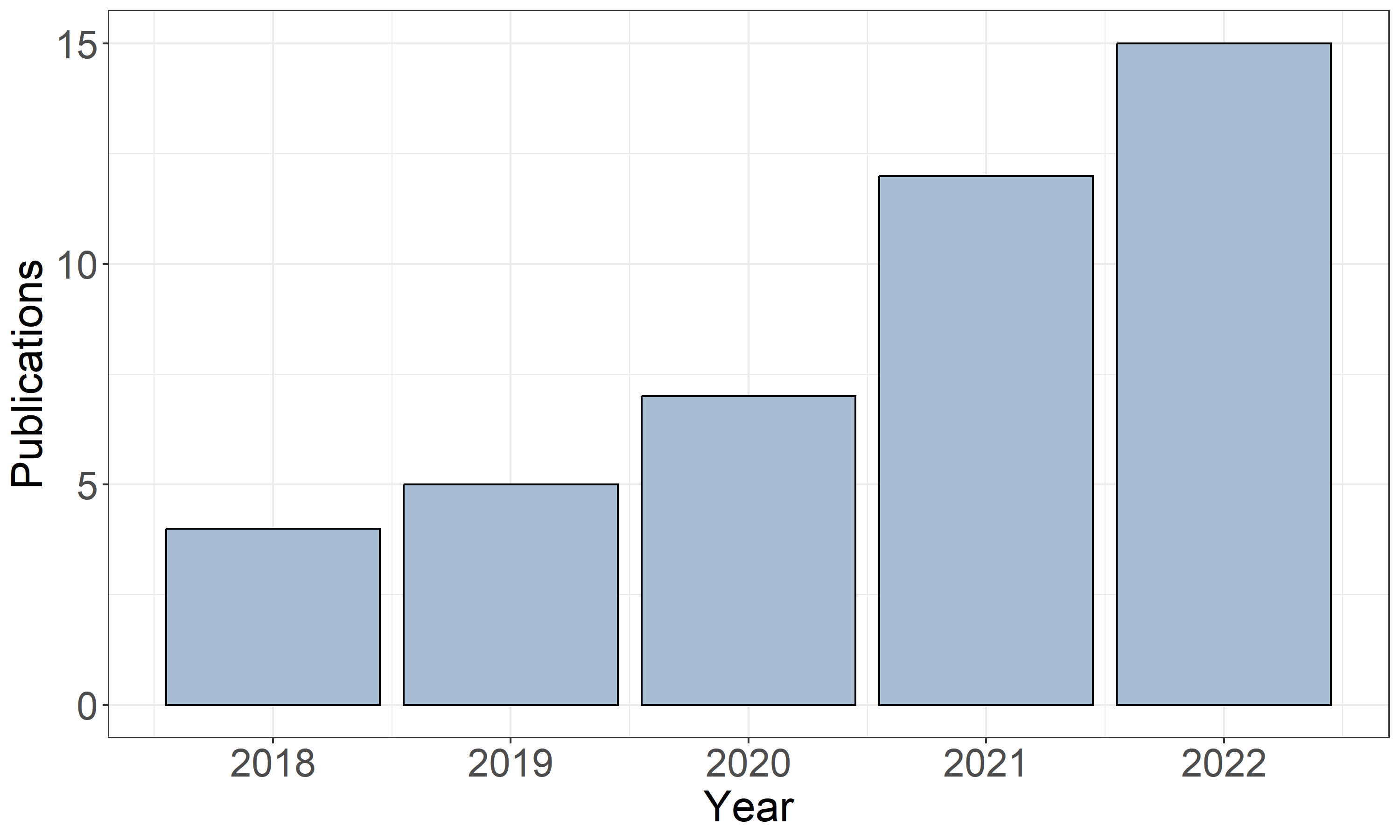}
    \caption{The number of publications over five years (between 2018-2022) in the leading economics journals that use ML. The data includes articles from the following ten journals: American Economic Review (AER), Econometrica, Journal of Economic Perspectives (JEP), Journal of Monetary Economics (JME), Journal of Political Economy (JPE), Journal of Econometrics (JoE), Quarterly Journal of Economics (QJE), Review of Economic Studies (RES), American Economic Journal (AJE): Macroeconomics and Microeconomics. The relevant papers are identified using the following search terms: Machine learning, Ensemble learning, Deep learning,  Statistical learning, Reinforcement learning, and Natural language processing.}
    \label{fig:papers}
\end{figure}

The number of academic publications that use ML tools has increased significantly in the last few years, as depicted in~\autoref{fig:papers}, which shows the number of articles published in ten leading economics journals that use ML. This trend is expected to continue as researchers explore new ways to apply ML techniques to a wide range of economic problems. Nevertheless, the suitability and  applicability of these tools is not widely understood among economists and data scientists.
To bridge this gap, in this article, we provide a curated review of selected papers published in prominent economics journals that employ ML tools. The objective of this review is to assist economists interested in leveraging ML tools for their research and analysis, as well as data scientists seeking to apply their skills to economic applications.\footnote{The article is motivated from the American Economic Association’s (AEA) continuing education session on Machine Learning and Big Data at the 2023 ASSA annual meeting~\cite{AEA2023}.}

The article aims to showcase the potential benefits of utilizing ML in economics research and policy analysis, while also offering suggestions on where, what and how to effectively apply ML models. It should be noted that the article takes a suggestive approach, rather than an explanatory one, with a particular focus on supervised learning methods commonly used for prediction problems such as regression and classification.
The article is organized into three main sections, each focusing on one key question: 1) When ML is used in economics? 2) What ML models are commonly preferred?, and 3) How ML is used for economic applications?  Finally, we briefly discuss the limitations of machine learning in its current state.

The key lessons of the review are summarized as follows: 
First, ML models are used to process nontraditional and unstructured data such as text and images, to capture strong nonlinearity that is difficult to capture using traditional econometric models, and to improve prediction accuracy, extract new information, or automate feature extraction when dealing with large but traditional datasets. ML tools are probably not useful for cases with small data complexity, and the traditional econometric models will likely to suffice. 

Second, the choice of ML model depends on the type of application and underlying data characteristics. For example, when conducting textual analysis, the latent Dirichlet allocation (LDA), which is an ML algorithm for probabilistic topic modelling, is commonly preferred. However, deep learning models such as Transformers are also employed for handling large text or audio data. Convolutional neural network models such as ConvNext are preferred for image or video datasets. Ensemble learning models are commonly employed for traditional datasets, and causal ML models are utilized when analysis focuses on causal inference.

Third, the usefulness of ML models can be greatly improved by tailoring them to the specific application, especially when dealing with large and complex but traditional data. This approach is more effective than using off-the-shelf tools. Moreover, pre-trained models with transfer learning can be advantageous when dealing with nontraditional but limited data and deep learning.\footnote{See~\autoref{apn:1} for more details on LDA model, \autoref{apn:2} for Transformer model, \autoref{apn:3} for ConvNext model, \autoref{apn:4} for ensemble learning models, and \autoref{apn:5} for transfer learning.}

This review highlights the potential benefits of using new tools provided by ML in various areas of economics research, but also acknowledges the challenges that needed to overcome to effectively use it for economic analysis and research. For instance, ML models require large amounts of data and ample computational resources, which can be a limitation for some researchers because it can be difficult to obtain high-quality data, and data may be incomplete or biased. Additionally, ML models are prone to overfitting and can be challenging to interpret, which limits their utility. Moreover, most ML models do not have standard errors, and other statistical properties have not yet been well-defined, which can make it difficult to draw conclusions from the results. Therefore, caution is recommended when using ML models. 
Despite these limitations, our review suggests, ML is successfully employed alongside traditional ecnometrics tools to advance our understanding of economic systems. By combining the strengths of both fields, researchers can improve the accuracy and reliability of economic analyses to better inform policy decisions. 

\section{When ML is used in economics?}
The literature suggest that the following three cases where the ML models could add value to economic research and analysis.
\begin{itemize}
    \item To process non-traditional data, such as images, texts, audios, and videos.
    \item To capture nonlinearity which is difficult to capture using traditional models.
    \item To process traditional data at scale to improve prediction accuracy, extract new information, or automate feature extraction.
\end{itemize}

For instance, article~\cite{varian2014big} suggest that the big data, due to its sheer size and complexity, may require more powerful manipulation tools, which ML can offer. Also, we may have more potential predictors (or features) than appropriate for estimation in some cases, so we need to make some variable selections, where ML can help. Lastly, large datasets allow for more flexible relationships than simple linear models can capture. ML techniques are handy in those cases due to their ability to model intricate and nonlinear relationships potentially offering new insights. 

Similarly, article~\cite{mullainathan2017machine} argue that ML not only provides new tools but also solves a different problem. They assert ML's success is largely due to its ability to discover the complex structure that was not specified in advance. They suggest applying ML to economics requires finding relevant tasks, for instance, where the focus is on increasing prediction accuracy or uncovering generalizable patterns from complex datasets. 
Also, article~\cite{athey2019machine} point that the methods developed in the ML have been particularly successful in big data settings, where we observe information on a large number of units, many pieces of information on each unit, or both. The authors suggest that for using ML tools for economics research and analysis, researchers should clearly articulate their goals and why certain properties of ML algorithms may or may not be important.

\subsection{ML is used for processing non-traditional data.}
Non-traditional datasets, such as images, text, and audio, can be difficult to process using traditional econometric models. In such cases, ML models can be used to extract valuable information that can be incorporated into traditional models to address economic questions.

For instance, article~\cite{hansen2018transparency} published in QJE, the authors assess how transparency, a key feature of central bank design, affects monetary policy makers’ deliberations, using an ML algorithm for probabilistic topic modelling. 
Similarly, article~\cite{larsen2021news} published at JME, they use a large news corpus and ML algorithms to investigate the role played by the media in the expectations formation process of households, and in~\cite{angelico2022can} published in JoE, uses the twitter data and ML model to measure inflation expectation.  
Likewise, artciel~\cite{henderson2012measuring} published in AER, the authors use satellite data to measure GDP growth at the sub and supranational regions, in~\cite{naik2016cities} also published in AER, the authors employed a computer vision algorithm that measures the perceived safety of streetscapes, and how strongly it is correlated with population density and household income, and in~\cite{gorodnichenko2023voice} published in AER, the authors use deep learning  to detect emotions embedded in press conferences after the Federal Open Market Committee meeting and examine the influence of the detected emotions on financial markets.

\subsection{ML is used for capturing strong nonlinearity}
The ML could be useful if the data and application contain strong nonlinearity, which is hard to capture using traditional approaches. For instance, in the article~\cite{kleinberg2018human}, published in QJE, the authors evaluate whether ML models can help improve judges' decisions on bail or no bail. Although the outcome is binary, this is a highly complex problem that demands processing complex data to make prudent decisions.

Similarly, in the article~\cite{maliar2021deep}, published in JME, the authors use ML to solve dynamic economic models by casting them into nonlinear regression equations. Here ML is used to deal with multicollinearity and to perform the model reduction.
Likewise, article~\cite{mullainathan2022diagnosing}, published in QJE, uses ML to test how effective physicians are at diagnosing heart attacks. Using a large and complex dataset available at the time of the physician's decision, they estimate the model to predict the outcome of testing to uncover potential sources of error in decisions.

\subsection{ML is used for processing traditional data at scale to improve prediction accuracy or extract new information}
ML could be useful for processing large and complex but traditional data sets with many variables. In such cases, the ML models can help to 1. improve prediction accuracy, 2. extract new information or 3. automate feature extraction. For instance, in the article~\cite{bianchi2022belief}, published in AER, the authors combine a data-rich environment with an ML model to provide new estimates of time-varying expectational errors embedded in survey responses and show that the ML can be productively deployed to correct errors in human judgment and improve predictive accuracy. 

Similarly, in the article~\cite{bandiera2020ceo}, published in JPE, the authors measure CEO behaviour types using high-frequency, high-dimensional diary data and an ML algorithm.
In~\cite{farbmacher2020explainable}, published in JoE, the author uses ML for fraud detection in insurance claims using unstructured data comprising inputs of varying lengths and variables with many categories. They argue that ML alleviates these challenges that are otherwise hard for traditional  methods. Likewise, in~\cite{dobbie2021measuring} published in RES, the authors suggest that the accuracy of  ML-based data driven decision rule for consumer lending could be more accurate than examiner-based decisions.

\begin{figure}[h!]
    \centering
    \includegraphics[width=0.57\textwidth]{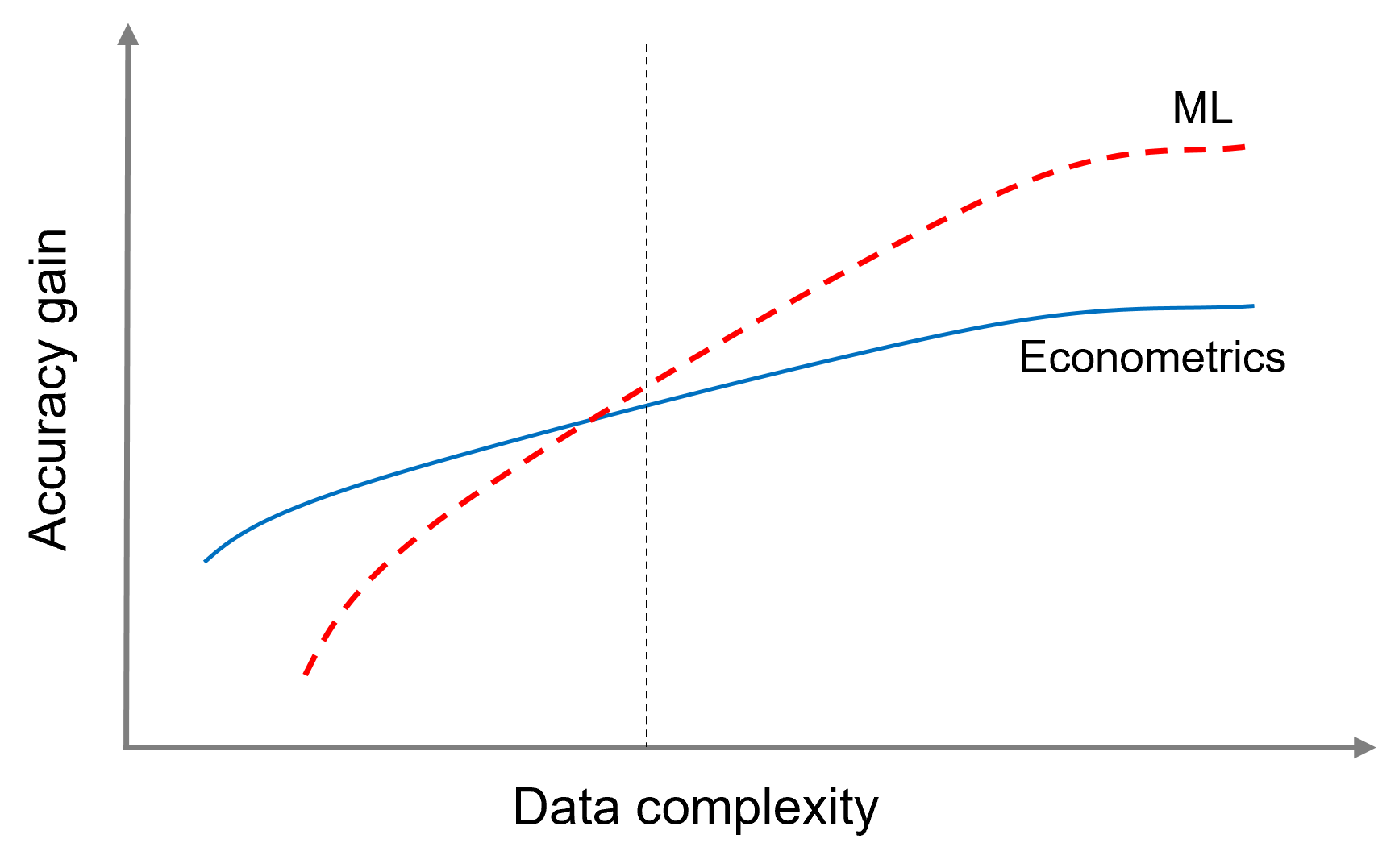}
    \caption{Schematic diagram representing the relative merits of ML and traditional econometric methods. The plot is adapted from~\cite{AEA2023,harding2018big}.}
    \label{fig:my_label}
\end{figure}

ML is probably not useful for the cases where data complexity (which could be related to shape, size, collinearity, nonlinearity, etc.) is small, and traditional econometric models would likely suffice. However, if the data complexity increases, i.e., when dealing with big data, the value added by ML models could be higher after a certain threshold, as shown by the dotted line in \autoref{fig:my_label}.

\section{What ML models are commonly preferred?}\label{sec:methods}

With a variety of ML tools available, different models are better suited for different types of applications and data characteristics. This section will discuss which models are most effective for a given type of economic application.

\subsection{Deep learning models are used when dealing with nontraditional data}

Natural language processing (or NLP) primarily relies upon processing textual data and has many applications in economics. For instance, it could be used for topic modelling or sentiment analysis. The model of choice for the topic modelling to quantify text is LDA proposed in~\cite{blei2003latent}. It is an ML algorithm for probabilistic topic modelling that decomposes documents in terms of the fraction of time spent covering a variety of topics~\cite{hansen2018transparency,larsen2021news}.

An review of the use of text as data and ML methods for various economics applications is presented in~\cite{gentzkow2019text}. 
The use of deep learning models for NLP is evolving rapidly, and various large language models (LLMs) could be used to process text data; However, \textit{transformer} models~\cite{vaswani2017attention} are proven to be more useful to efficiently extract useful information from the textual data~\cite{AEA2023}. For instance, in~\cite{gorodnichenko2023voice}, they use transformer to detect emotions embedded in press conferences after the Federal Open Market Committee meeting for sentiment analysis. Moreover, almost all large general-purpose LLMs, including GPT-3 and chatGPT, are trained using Generative Pre-trained Transformer~\cite{brown2020language}. 

An interesting application of computer vision models in economics is using a broad set of satellite images or remote sensing data for analysis. For instance, in the article~\cite{henderson2012measuring}, the authors use satellite data to measure GDP growth at the sub and supranational regions. Similarly, in~\cite{xie2016transfer}, using satellite data and machine learning, they develop high-resolution predictors of household income, wealth, and poverty rates in five African countries. 

A review of the literature, presenting opportunities and challenges in using satellite data and ML in economics, is documented in~\cite{donaldson2016view}. It concludes that such models have the potential to perform economic analysis at spatial and temporal frequencies that are an order of magnitude higher than those that are commonly available. 
Various deep learning models can be employed to extract useful information from images, but the \textit{ConvNext} model~\cite{liu2022convnet} is evolving to be more successful in efficiently processing image data sets~\cite{AEA2023}. However, the transformers can also be effectively employed for image processing~\cite{chen2020generative}.

\subsection{Ensemble learning models are used when dealing with traditional data}

Ensemble learning models could be useful if the data size is small but includes many features and if there is collinearity or nonlinearity, which is hard to model. 
For instance, the article~\cite{mullainathan2017machine} compares the performance of different ML models in predicting house prices and demonstrates that the nonparametric ML algorithms, such as random forests, can do significantly better than ordinary least squares, even at moderate sample sizes and
with a limited number of covariates. 

Similarly, article~\cite{mullainathan2022diagnosing} use ensemble learning models which combine gradient-boosted trees and LASSO to study physicians' decision-making to uncover the potential sources of errors. 
Also, in~\cite{athey2019generalized}, they propose generalized random forests, a method for nonparametric statistical estimation based on random forests. This can be used for three statistical tasks: nonparametric quantile regression, conditional average partial effect estimation and heterogeneous treatment effect estimation via instrumental variables. Moreover, the ensemble learning models are popular in macroeconomic prediction, as demonstrated by many articles~\cite{richardson2020nowcasting,yoon2021forecasting,chapman2021macroeconomic,goulet2022machine,bluwstein2023credit}). 

\subsection{Causal ML models are used when the focus is on causal inference}
The causal ML could be helpful when the primary objective is to make causal inferences, but the dataset is big and complex. For instance, in~\cite{wager2018estimation}, the authors develop a nonparametric causal forest for estimating heterogeneous treatment effects that extends
the random forest, an ML algorithm. They demonstrate that any type of random forest, including classification and regression forests, can be used to provide valid statistical inferences. In experimenting with these models, they find causal forests to be substantially more powerful than classical methods---especially in the presence of irrelevant covariates. 

Similarly, the article~\cite{athey2016recursive} use ML for estimating heterogeneity in causal effects in experimental and observational studies. Their approach is tailored for applications where there may be many attributes of a unit relative to the number of units observed and where the functional form of the relationship between treatment effects and the attributes is unknown. It enables the construction of valid confidence intervals for treatment effects, even with many covariates relative to the sample size, and without sparsity assumptions. The applicability of these methods is demonstrated in~\cite{davis2017using} for predicting treatment heterogeneity for summer jobs application.

\section{How ML is used for economic applications}

Pre-trained models and transfer learning are recommended, especially while using deep-learning models~\cite{AEA2023}. Deep Learning based approaches are state-of-the-art for processing non-traditional data; However, there are many difficulties in using them for economic applications. For instance, large data and ample computational resources are often necessary to train these models. Both of which are scarce resources in many economic applications. Moreover, these models are notoriously convoluted, and many economics researchers would benefit from using these methods but lack the technical background to implement them from scratch (\cite{shen2021layoutparser}). Therefore, transfer learning, i.e., large models pre-trained on a similar application (for instance, large language or computer vision models), could be adapted for similar economic applications. 

Of-the-shelf ensemble learning models are useful when dealing with panel data with strong collinearity or nonlinearity, but it is recommended to adapt these models to suit the task. For instance, in~\cite{athey2019generalized}, they adapt the popular random forests algorithm, which then can be used for nonparametric quantile regression or estimating average treatment effects. Similarly, in~\cite{farbmacher2020explainable} propose an explainable attention network for fraud detection in claims management by adapting a standard neural network and demonstrate that the adapted model performs better than off-the-shelf. 
Likewise, in~\cite{goulet2020macroeconomy}, the author proposed macroeconomic random forest, an algorithm adapting the canonical ML tool. They show that the adapted model forecasts better than off-the-shelf ML algorithms and traditional econometric approaches, and it can be interpreted.

There are many other examples where the model or the procedure to train the model is adapted to improve the performance. For instance, \cite{athey2016recursive} adopt the standard cross-validation approach, which then enables the construction of valid confidence intervals for treatment effects, even with many covariates relative to the sample size. Similarly, a variation of cross-validation approaches is proposed in~\cite{chapman2021macroeconomic} to improve the prediction performance of the macroeconomic nowcasting model during economic crisis periods. 

A few other practical recommendations discussed in the AEA's course are: 1. Bigger, both in terms of data and model size, is better when using deep learning models for processing non-traditional data. 2. Due to its speed and community support, the \textit{Python} programming language is preferred over languages for applied ML. 3. When training large ML models, it is suggested to use \textit{Unix}-based operating systems over windows systems.

\section{Other emerging applications}
Other types of ML approaches, such as unsupervised learning and reinforcement learning, are yet to make a notable impact in economics research. However, there are some initial applications of these approaches in the literature. For instance, in~\cite{gu2021autoencoder} uses an unsupervised dimension reduction model called autoencoder neural network for asset pricing models. Similarly, in~\cite{triepels2017anomaly} autoencoder-based unsupervised model is used for anomaly detection in high-value payments systems. Also in~\cite{decarolis2021mad} using data on nearly 40 million Google keyword auctions and unsupervised machine learning algorithms to cluster keywords into thematic groups serving as relevant markets. 

Reinforcement learning (RL) models could be employed to model complex strategic decisions arising in many economic applications. For instance, in~\cite{castro2020estimating}, the authors use RL to estimate optimal decision rules of banks interacting in high-value payment systems. Similarly, in~\cite{chen2021deep}, deep RL is used to solve dynamic stochastic general equilibrium models for adaptive learning at the interaction of monetary and fiscal policy, and \cite{hinterlang2021optimal}, the authors use RL to optimize monetary policy decisions. Likewise, in~\cite{zheng2022ai}, the authors use a RL approach to learn dynamic tax policies. 

\section{Limitations}
The key limitations of ML for economics research and analysis are outlined below:
\begin{itemize}
    \item Large data sets and ample computational resources are often necessary to train ML models---especially the deep learning models.
    \item The ML models, owing to their flexibility, are easy to overfit, and their complexity makes them hard to interpret---which is crucial for many economic applications. 
    \item Most ML models have no standard errors and asymptotic properties---which could be essential for many economic applications.
    \item ML models can be biased if the data used to train these models is of low quality and biased. 
\end{itemize}

The literature is evolving to address these challenges; however, some are hard and could take longer to mitigate. For instance, we have limited data in many economic applications, which restricts the applicability of large ML models.  
This could be potentially mitigated in certain applications as the economy becomes more digital, allowing us to gather more data at a much higher frequency than traditional economics data sets. 

The interpretability or explainability of models is another major challenge in using ML in economics. The researchers are making progress toward overcoming these challenges. For instance, one approach recently developed to mitigate interpretability issues is by using the Shapley-value-based
methodologies such as developed in~\cite{lundberg2017unified}, and \cite{buckmann2021opening}. These methods are useful for macroeconomic prediction models in~\cite{buckmann2021opening,chapman2021macroeconomic,liuinterpreting2022,bluwstein2023credit}. However, note that, although such methods are based on game theory, they do not provide any optimal statistical criterion, and asymptotics for such approaches are not available yet. To overcome that, for instance, in the recent papers~\cite{babii2021machine,babii2022machine}, the authors propose ML-based mixed data sampling and develop the asymptotics in the context of linear regularized regressions. However, much progress needs to be made to use such asymptotic analysis for popular nonlinear ML approaches. 

\section{Conclusions}\label{sec:conclusion}

This concise review highlights that ML is increasingly used for economics research and policy analysis, particularly for analyzing non-traditional data, capturing nonlinearity, and improving prediction accuracy. Importantly, ML can complement traditional econometric tools by identifying complex relationships and patterns in data that can be incorporated into econometric models. As the digital economy and economic data continue to grow in complexity, ML remains a valuable tool for economic analysis. However, a few limitations need to be addressed to improve the utility of ML models, and the literature is progressing toward mitigating those challenges. 

\begin{figure}[h!]
     \centering
     \begin{subfigure}[b]{0.75\textwidth}
         \centering
          \includegraphics[width=\textwidth]{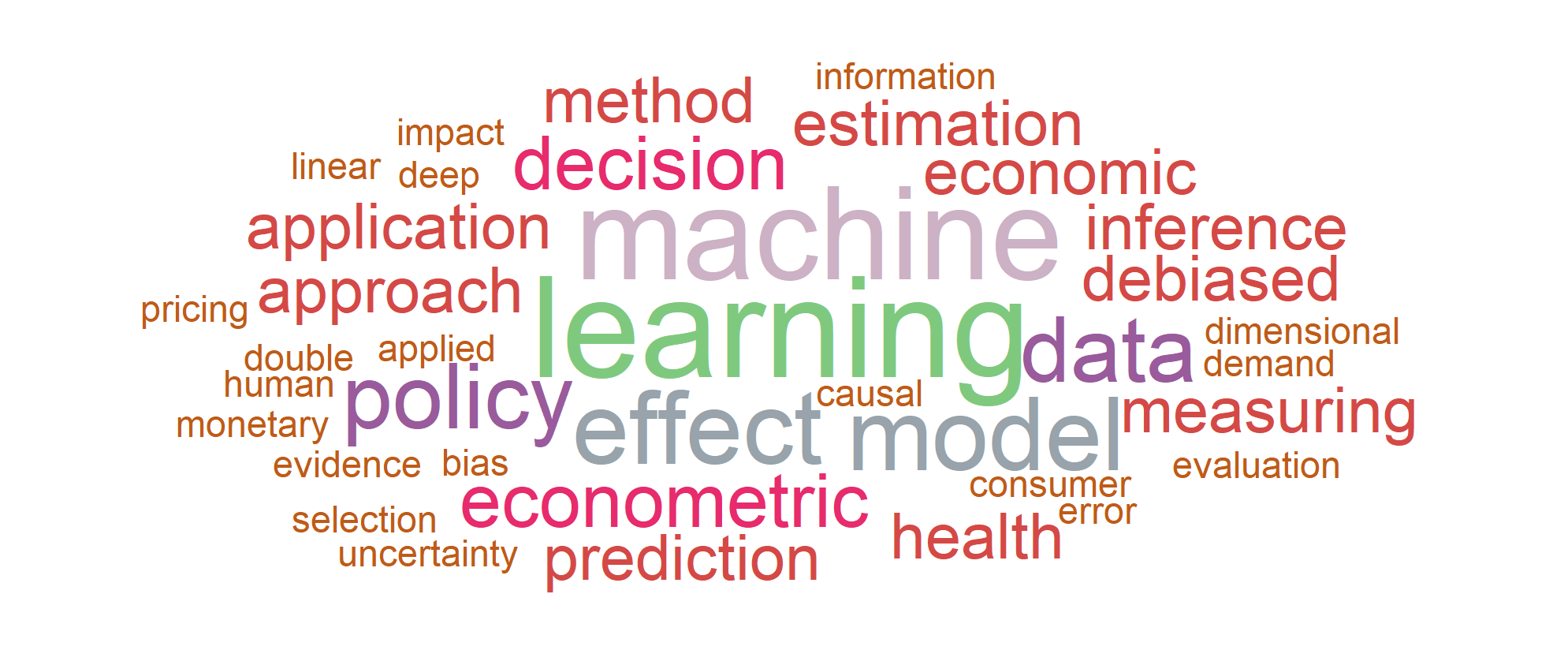}
         \caption{Word cloud for titles}
         \label{fig:y equals x}
     \end{subfigure}
     \hfill
     \begin{subfigure}[b]{0.75\textwidth}
         \centering
         \includegraphics[width=\textwidth]{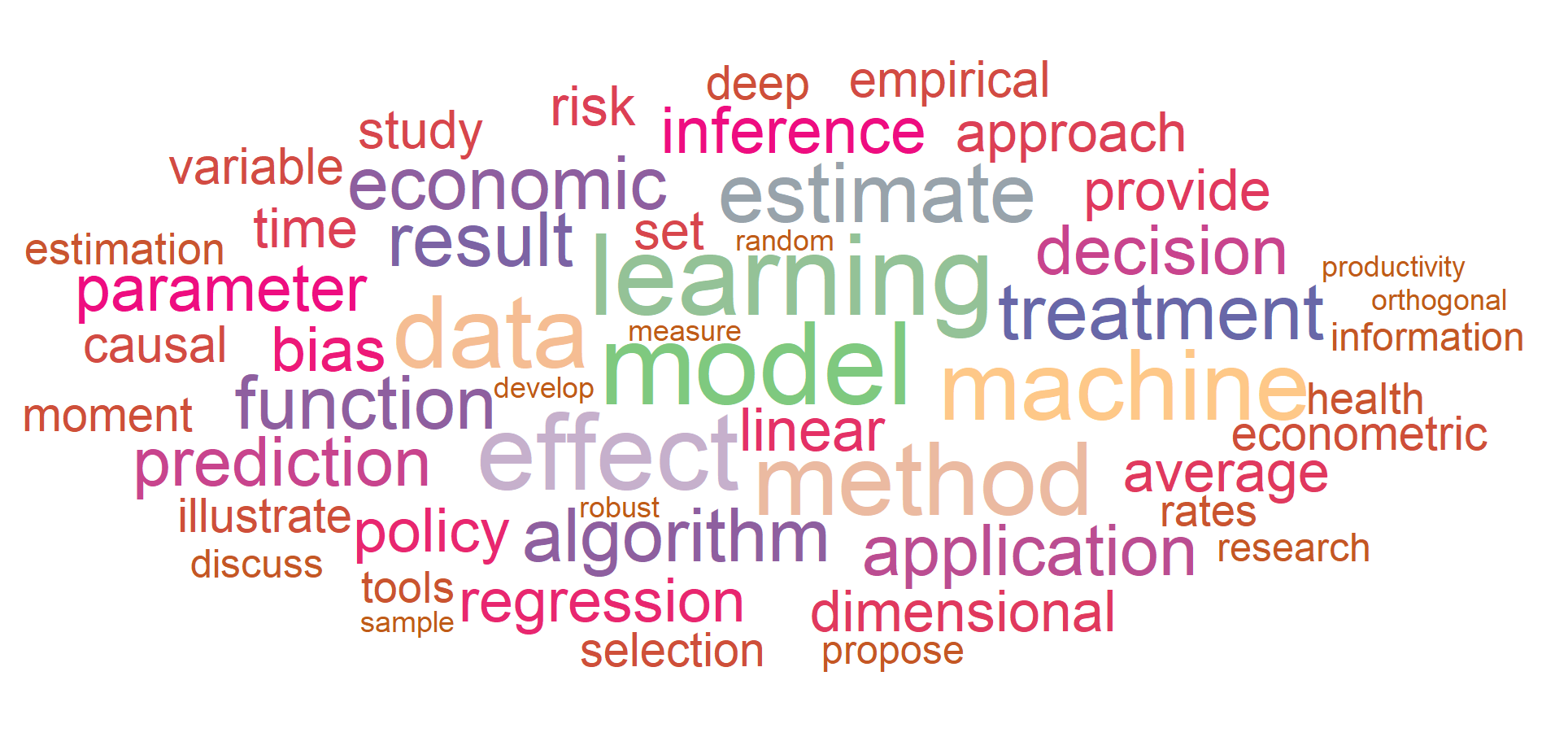}
         \caption{Word cloud for the abstracts}
         \label{fig:three sin x}
     \end{subfigure}
     \hfill
        \caption{Our dataset of articles collected from prominent economics journals using ML is visualized through word clouds of their titles and abstracts.}
        \label{fig:word_clouds}
\end{figure}

Lastly, in~\autoref{fig:word_clouds}, we present the word clouds generated from the titles and abstracts of the articles in our dataset. These word clouds illustrate the frequency of certain terms, with larger font sizes indicating more frequent usage. For example, as shown in the word clouds, the terms ``machine" and ``learning" are prominently featured in both titles and abstracts, highlighting its relevance in those articles. This is followed by words such as ``data," ``effect," and ``decision".  

\singlespacing
\bibliographystyle{unsrt}  
\bibliography{bibliography}


\appendix
\singlespacing
\section{Latent Dirichlet Allocation}\label{apn:1}
Latent Dirichlet Allocation is a probabilistic ML model commonly used for topic modelling. It assumes that each text document in a corpus contains a mixture of various topics which reside within a latent layer and that each word in a document is associated with one of these topics. The model infers those topics based on the distribution of words across the entire corpus. The output of the model is a set of topic probabilities for each document and a set of word probabilities for each topic. It has many practical applications, including text classification, information retrieval, and social network data analysis. Refer to~\cite{blei2003latent} for more details on the LDA model and its formulation. 

\section{Transformers}\label{apn:2}

The Transformers are a deep learning model architecture commonly used in text and image processing tasks. The key feature of transformers is self-attention mechanism to process sequential input data, such as words in a sentence. This self-attention allows the model to identify the most relevant parts of the input sequence for each output. Refer to~\cite{vaswani2017attention} for more details of the model.

Generally, the transformer architecture comprises an encoder and a decoder, consisting of multiple self-attention layers and feed-forward neural networks. The encoder processes the input sequence, such as a sentence in one language, and produces a sequence of context vectors. The decoder uses these context vectors to generate a sequence of outputs, such as translating into another language.
The key benefit of transformer models is their ability to handle long-range dependencies in input sequences, making them particularly effective for NLP tasks that require understanding the context of words or phrases within a longer sentence or paragraph.

\section{ConvNext}\label{apn:3}

The ConvNeXT is a convolutional neural network (CNN) model inspired by the design of the type of transformers model architecture. It is a deep learning model commonly used for processing image and video data for various tasks, for instance, object detection, image classification, and facial recognition. Refer to~\cite{liu2022convnet} for more details on the model and its formulation.

Generally, CNN models comprise multiple layers sandwiched between input and output layers. It includes convolutional, pooling, and fully connected layers. In the convolutional layers, the model performs a set of mathematical operations called convolution on the input image to extract high-level features such as edges, corners, and textures. The pooling layers are used to downsample the convolutional layers' output, reducing the feature maps' size for the subsequent layers. Finally, the fully connected layers are used to classify the image based on the extracted features. One of the key benefits of CNNs is their ability to learn spatial hierarchies of features, i.e., the model can identify complex patterns and objects in images by learning to recognize simpler features first and gradually build up to more complex ones.

\section{Ensemble Learning}\label{apn:4}

Ensemble learning is a popular ML technique that combines multiple individual models, called base models or weak learners, to improve the accuracy and robustness of the overall prediction. Each base model is trained on randomly sampled subsets of the training data, and their predictions are then combined in some way to produce a final prediction. Ensemble learning using decision trees (DTs) is popular among others. By combining multiple DTs trained on different subsets of the data and using different parameters, ensemble learning can capture a broader range of patterns and complex relationships in the data to produce more accurate and robust predictions. 

Bagging and boosting are two commonly used ensemble learning approaches. The bagging involves creating multiple decision trees, each trained on a randomly sampled subset of the training data. The final prediction is made by combining the predictions of all the individual decision trees, such as by taking the average or majority vote of the individual tree predictions. Random Forest~\cite{breiman2001random} is a popular example of this approach. 
On the other hand, boosting involves training DTs sequentially, with each new tree attempting to correct the errors of the previous tree by using the modified version of the original training data. The final prediction is made by combining the predictions of all the individual decision trees, with greater weight given to the predictions of the more accurate trees. A popular example is gradient boosting ~\cite{natekin2013gradient}. The advanced version of boosting methods such as XGBoost~\cite{chen2015xgboost} and LightGBM~\cite{ke2017lightgbm} are popular among others.
Check out~\cite {sagi2018ensemble}, a review article that details ensemble learning, its types,  applications and limitations.

\section{Transfer Learning}\label{apn:5}
Transfer learning is an ML technique that involves leveraging knowledge gained from training a model on one task to improve the performance of a model on a different but related task. In transfer learning, a pre-trained model is used as a starting point for a new task, hoping that the knowledge and patterns learned by the model in the pre-training stage will be useful for the new task. The pre-trained model can be trained on a similar or different dataset.

Transfer learning is a powerful technique particularly useful in deep learning because it can help improve the accuracy of models and reduce the amount of data required for training. By using a pre-trained model as a starting point for a new task, the model can leverage the knowledge and patterns learned from the previous task to improve its performance on the new task. This can lead to faster convergence, better generalization, and improved accuracy, especially in situations where the new task has limited data available for training. 
Generally, it is effective in cases for large models with millions of parameters and requires significant amounts of data for training. Transfer learning has been successfully applied in a wide range of tasks related to nontraditional data processing. Check out~\cite{xie2016transfer}, a review article which provides further details on transfer learning, its types, applications and limitations.

\end{document}